\documentclass[12pt]{article}
\usepackage{times,amsmath,amsfonts,amsthm,amssymb,eucal}
\usepackage[utf8]{inputenc}
\usepackage{algorithm}
\usepackage{algorithmic}
\usepackage{graphicx,ifthen}
\usepackage{wrapfig}
\usepackage{latexsym}
\usepackage{color}
\usepackage{mathrsfs}
\usepackage{multirow}
\usepackage{bm}
\usepackage{bbm}
\DeclareGraphicsExtensions{.gif,.pdf,.png,.jpg,.tiff}
\begin{document}
\renewcommand{\baselinestretch}{1.54}
\def\Quote{\begin{quotation}\normalfont\small}
\def\EndQuote{\end{quotation}\rm}
\def\BigHeading{\bfseries\Large}\def\MediumHeading{\bfseries\large}
\def\bct{\begin{center}}
\def\ect{\end{center}}
\font\BigCaps=cmcsc9 scaled \magstep 1
\font\BigSlant=cmsl10    scaled \magstep 1
\def\lbk{\linebreak}
\def\Report{Inference in High-dimensional GWAS}
\def\Author{}
\pagestyle{myheadings}
\markboth{\Author}{\Report}
\thispagestyle{empty}
\bct{\BigHeading Variance Estimation and Confidence Intervals from
High-dimensional Genome-wide Association Studies Through
Misspecified Mixed Model Analysis}
\\\vskip10pt
\BigCaps {Cecilia ${\rm Dao}^{1}$, Jiming ${\rm Jiang}^{2}$, Debashis
${\rm Paul}^{2}$ and Hongyu ${\rm Zhao}^{1}$\lbk
\BigSlant Yale University, ${\rm USA}^{1}$ and University of California,
Davis, ${\rm USA}^{2}$}
\ect
\Quote
\vskip-5pt
We study variance estimation and associated confidence
intervals for parameters characterizing genetic effects from genome-wide association studies
(GWAS) misspecified mixed model analysis. Previous studies have shown
that, in spite of the model misspecification, certain quantities of genetic
interests are estimable, and consistent estimators of these quantities can
be obtained using the restricted maximum likelihood (REML) method
under a misspecified linear mixed model. However, the asymptotic
variance of such a REML estimator is complicated and not ready to be
implemented for practical use. In this paper, we develop practical and computationally convenient methods for estimating
such asymptotic variances and constructing the associated confidence
intervals. Performance of the proposed methods is evaluated empirically
based on Monte-Carlo simulations and real-data application.

\EndQuote

\section{Introduction}

\hspace{4mm}
Genome-wide association studies (GWASs) have proved successful by scanning the genome for genetic variations, e.g., single nucleotide polymorphisms (SNPs), that are associated with disease status and traits across study subjects. Tens of thousands of SNPs have been identified to be associated with various diseases and traits. For a review of the GWAS remarkable discoveries, see Visscher {\it et al.}  (2017). Researchers can use GWAS results to further medical research, such as to determine a person’s risk of developing a disease or treat/prevent the disease. It is well known that genetic factors may account substantially for disease risk or various traits, which can be quantified as heritability. Historically, heritability was inferred from resemblance among different degrees of related individuals (e.g., twin studies) without studying specific genetic variations, but today there is an emerging interest in quantifying how much variation can be accounted for from GWAS data due to the recent development of efficient genotyping and sequencing technology and the success of the GWAS strategy. 
However, when GWAS significant variants were considered, they only explained a small fraction of the genetic component of the phenotypes. The gap between the phenotypic variance explained by significant GWAS results and that estimated from classical heritability methods is known as the ``missing heritability problem." 

More precisely, the problem refers to the concept that SNPs that are significant in GWASs cannot fully account for heritability of many diseases and traits. One explanation for missing heritability is that many SNPs jointly affect the phenotype, and SNPs with smaller effects that have not been identified may contribute to heritability as well. To address this issue, Yang {\it et al.}  (2010) used an approach involving linear mixed models (LMMs) to show that a large proportion of heritability is not missing but rather captured by SNPs with weak effects that do not reach genome-wide significance level. The general idea is to use an LMM to treat the effects of all SNPs as random effects rather than relying on single-SNP association analysis. This approach has been widely used for heritability estimation in the genetics community via the GCTA software (Yang {\it et al.} 2011).

In an attempt to make the modelling more accurate, others have proposed extensions of this LMM approach. For instance, Heckerman {\it et al.}  (2016) proposed to add an environmental random effect (along with a genomic random effect) in the LMM to reduce heritability inflation, and Zhou {\it et al.}  (2013) proposed to use a hybrid of LMM and spare regression models to learn the true genetic architecture from the data to estimate heritability. To improve heritability estimation compared to GCTA, Speed {\it et al.} (2017)  developed the LDAK model to factor in minor allele frequency (MAF), linkage disequilibrium (LD), and genotype certainty. Speed {\it et al.} (2020) extended their LDAK model to handle more complex heritabiltiy models by proposing an approximate model likelihood to be computed by GWAS summary statistics. Comprehensive comparisons of heritability estimation methods [Yang, Manolio, {\it et al.} (2011), Yang {\it et al.} (2015), Speed {\it et al.} (2017), Speed {\it et al.} (2012), Zaitlen {\it et al.} (2013), Bulik-Sullivan {\it et al.} (2015)] can be found in Evans {\it et al.} (2018). Zhu and Zhou (2020) also provides a review of statistical methods for heritability estimation.

Consider an LMM which can be expressed as
\begin{equation}
y=X\beta + \tilde{Z}\alpha + \epsilon,
\end{equation}
where $y$ is an $n \times 1$ vector of observations; $X$ is an $n \times q$ matrix of known covariates;
$\beta$ is a $q \times 1$ vector of unknown regression coefficients (the fixed effects); and $\tilde{Z} = p^{-1/2}Z$, where $Z$ is an $n \times p$ matrix whose entries are random variables. Furthermore, $\alpha$ is a $p \times 1$ vector of random effects that are distributed as $N(0,\sigma^2_\alpha I_p)$, $I_p$ being the $p$-dimensional identity matrix, and $\epsilon$ is an $n \times 1$ vector of errors that
is distributed as $N(0,\sigma^2_\epsilon I_n)$, and $\alpha$, $\epsilon$ and $Z$ are independent. The heritability parameter is defined as $h^2 = \sigma^2_{\alpha}/(\sigma^2_{\alpha} + \sigma^2_{\epsilon})$.

The LMM (1) is the model used by Yang {\it et al.}  (2010) where it is assumed the effects of all the SNPs (random effects) are nonzero. The restricted maximum likelihood (REML) estimator of the heritability is given by $\hat{h}^2 = \hat{\sigma}^2_{\alpha}/(\hat{\sigma}^2_{\alpha} + \hat{\sigma}^2_{\epsilon})$, where the estimates of the variance components $\hat{\sigma}^2_{\alpha}$ and $\hat{\sigma}^2_{\epsilon}$ are based on the REML method [e.g., Jiang (2007), Section 1.3.2], which is implemented in the GCTA software.
In reality, however, only a subset of the SNPs are potentially nonzero. Specifically, we have $\alpha = \{\alpha'_{(1)}, 0'\}$, where $\alpha'_{(1)}$ is the vector of the first $m$ components of $\alpha$ $(1 \leq m \leq p)$, and $0$ is
the $(p-m)\times1$ vector of zeros. Correspondingly, we have $\tilde{Z} = [\tilde{Z}_{(1)} ; \tilde{Z}_{(2)}]$, where $Z_{(j)} = p^{-1/2}Z_{(j)}, j = 1, 2, Z_{(1)}$  is $n \times m$, and $Z_{(2)}$ is $n \times (p-m)$. Therefore, the true LMM can be expressed as
\begin{equation}
y=X\beta + \tilde{Z}_{(1)}\alpha_{(1)} + \epsilon,
\end{equation}
With respect to the true model (2), the assumed model (1) is misspecified. We call the latter a misspecified LMM, or mis-LMM. 

Jiang {\it et al.}  (2016) showed that even under a mis-LMM, $\hat{\sigma}_{\epsilon}^{2}$ and $\hat{h}^2$ are consistent by investigating the asymptotic behavior of the estimators as the sample size and the number of SNPs increase to infinity, such their ratio converges to a finite, nonzero constant. However, the asymptotic variances of the REML estimators have complex forms that are not ready to be implemented for practical use. This issue is important, from a practical point of view, because the asymptotic variance  is used to obtain the standard error of the estimator, and confidence interval for the associated parameter, in applications. The main goal of the current paper is to propose accurate estimators of the variance of $\hat{\sigma}_{\epsilon}^{2}$ and $\hat{h}^2$ along with confidence intervals that are robust even under the mis-LMM. The proposed variance estimators are derived based on asymptotic approximation; they have analytic expressions and are simple to use. Using the variance estimators and in results Jiang {\it et al.} (2016), we constructed approximate $100(1 - \alpha)\%$ confidence intervals for the associated parameters. We also considered a nonparametric approach to construct bootstrap confidence intervals. Particularly, let $F$ be the true distribution of $\theta(F) \in \{\sigma_{\epsilon}^{2}, h^2\}$, and let $\hat{\theta} \equiv \theta(\hat{F}) \in \{\hat{\sigma}_{\epsilon}^{2}, \hat{h}^2\}$ be the REML estimate of $\theta(F)$. Let $\hat{F^*}$ denote a bootstrap approximation to $\hat{F}$. Since the sampling distribution of $\theta(\hat{F})/\theta(F) \approx \theta(\hat{F^*})/\theta(\hat{F})$, we constructed an approximate $100(1 - \alpha)\%$ confidence interval for $\theta$ as $ \left(\frac{\hat{\theta}}{q^*_{1 - \alpha/2}},  \frac{\hat{\theta}}{ q^*_{\alpha/2}} \right)$, where $q^*_{t}$ is the $t$-th quantile of the bootstrap sampling distribution of $\theta(\hat{F^*})/\theta(\hat{F})$. Since the first method of confidence interval construction performed better based on empirical coverage probabilities, we present it in this paper. 

There have been studies in recent literature regarding uncertainty measures associated with the GCTA method of Yang {\it et al.} (2010). Kumar {\it et al.} (2016) showed that the confidence intervals produced by the GCTA method overwhelmingly underestimated the uncertainty in the variance component estimates even when the LMM model assumptions are satisfied. Also, see Lohr {\it et al.} (1997) and  Burch (2007) regarding inaccuracy of the confidence intervals. We carried out empirical studies regarding the performance of our proposed variance estimators and associated confidence intervals as well as those of GCTA variance estimator. In our simulation studies under misspecification, we found that the GCTA method is satisfactory in terms of variance estimators and associated confidence intervals. In fact, the two methods, our proposed method and GCTA method, performed similarly in our simulation studies. Note that our method is supported by the theory established in Jiang {\it et al.} (2016). Thus, in a way, our findings also restore confidence in GCTA in terms of measures of uncertainty even under misspecification.

In Section 2, we derive the variance estimators and associated confidence intervals. In Section 3 we demonstrate performance of the methods using simulation studies.  Section 4 contains a real data example using the UK Biobank data. Technical details are deferred to the Appendix.
\section{Derivation of variance estimators}
\hspace{4mm}
As noted, Jiang {\it et al.} (2016) showed that REML estimators of certain variance components of genetic interest are consistently estimable and asymptotically normal; however, the corresponding asymptotic variances do not have expressions suitable for implementation. Thus, our first objective is to derive (simple) estimators of those asymptotic variances. Let us begin with estimation of ${\rm var}(\hat{\sigma}_{\epsilon})$. By Jiang {\it et al.} (2016), we have the expression
\begin{eqnarray}
\hat{\sigma}_{\epsilon}^{2}&=&\frac{y'P_{\hat{\gamma}}^{2}y}{{\rm
tr}(P_{\hat{\gamma}})},
\end{eqnarray}
where $\hat{\gamma}=\hat{\sigma}^{2}_{\alpha}/\hat{\sigma}^{2}_{\epsilon}$. Some technical (see Subsection A.1 of the Appendix) derivations lead to the following approximation:
\begin{eqnarray}
\hat{\sigma}_{\epsilon}^{2}&\approx&\frac{{\rm E}(U_{\gamma,y})}{{\rm E}(U_{\gamma,y})-{\rm
E}(S_{\gamma,y})}\cdot\frac{y'P_{\gamma}^{2}y}{{\rm tr}(P_{\gamma})}+\frac{{\rm E}(U_{\gamma,y})}{{\rm E}(S_{\gamma,y})-{\rm
E}(U_{\gamma,y})}\cdot\frac{y'Q_{\gamma}y}{{\rm tr}(P_{\gamma}\tilde
{Z}\tilde{Z}')},
\end{eqnarray}
where $\gamma=\gamma_{*}$, which is the asymptotic limit of $\hat{\gamma}$ according to Jiang {\it et al.} (2016).
Denote the right side of (4) by $\tilde{\sigma}_{\epsilon}^{2}$, then, we have
\begin{eqnarray}
{\rm var}(\hat{\sigma}_{\epsilon}^{2})\approx{\rm var}(\tilde
{\sigma}_{\epsilon}^{2})={\rm E}\{{\rm var}(\tilde{\sigma}_{\epsilon}^{2}|
Z)\}+{\rm var}\{{\rm E}(\tilde{\sigma}_{\epsilon}^{2}|Z)\}.
\end{eqnarray}
It can be shown that the second term on the right side of (5) is of lower order than the first term; therefore, we have
\begin{eqnarray}
{\rm var}(\hat{\sigma}_{\epsilon}^{2})&\approx&{\rm  E}\{{\rm var}(\tilde
{\sigma}_{\epsilon}^{2}|Z)\}.
\end{eqnarray}
To obtain a further approximation, define
\begin{eqnarray*}
A&=&\frac{{\rm tr}(Q_{\gamma}\tilde{Z}\tilde{Z}')\{{\rm tr}(P_{\gamma}^{2})
{\rm tr}(Q_{\gamma}\tilde{Z}\tilde{Z}')-{\rm tr}^{2}(Q_{\gamma})\}}{{\rm
tr}^{2}(P_{\gamma}){\rm tr}^{2}(P_{\gamma}\tilde{Z}\tilde{Z}')},\\
B&=&\frac{{\rm tr}(Q_{\gamma}\tilde{Z}\tilde{Z}')}{{\rm tr}(P_{\gamma}\tilde
{Z}\tilde{Z}')}-\frac{{\rm tr}(Q_{\gamma})}{{\rm tr}(P_{\gamma})}.
\end{eqnarray*}
Then, it can be shown (see Subsection A.2 of the Appendix) that the right side of (6) can be approximated by
$2\sigma_{\epsilon}^{2}{\rm E}(A)/\{{\rm E}(B)\}^{2}$. Thus, in conclusion, we obtain the following estimator of ${\rm
var}(\hat{\sigma}_{\epsilon})$:
\begin{eqnarray}
\widehat{{\rm var}(\hat{\sigma}_{\epsilon}^{2})}&=&2\hat{\sigma}_{\epsilon}^{2}\frac{\hat{A}}{\hat{B}^{2}},
\end{eqnarray}
where $\hat{A}, \hat{B}$ are $A, B$ with $\gamma$ replaced by $\hat{\gamma}$, respectively.

Next, we consider estimation of ${\rm var}(\hat{\gamma})$. Using similar arguments (details omitted), it can be shown that
\begin{eqnarray}{\rm var}(\hat{\gamma})&\approx&\frac{{\rm E}(C)}{\{{\rm E}(B)\}^{2}},
\end{eqnarray}
where
\begin{eqnarray*}
C&=&{\rm tr}\left[\left\{\frac{P_{\gamma}}{{\rm tr}(P_{\gamma})}-\frac{P_{\gamma}\tilde{Z}\tilde{Z}'}{{\rm tr}(P_{\gamma}\tilde
{Z}\tilde{Z}')}\right\}^{2}\right].
\end{eqnarray*}
Thus, an estimator of ${\rm var}(\hat{\gamma})$ is given by
\begin{eqnarray}
\widehat{{\rm var}(\hat{\gamma})}&=&\frac{\hat{C}}{\hat{B}^{2}}.
\end{eqnarray}

The variance estimator (9) is used to obtain a variance estimator for $\hat{h}^{2}$, the REML estimator of the heritability, $h^{2}$. Using the expression $\hat{h}^{2}
=\hat{\gamma}/(1+\hat{\gamma})$, and the delta-method (e.g., Jiang 2010, sec. 4.2), we obtain
\begin{eqnarray}
{\rm var}(\hat{h}^{2})&\approx&\frac{{\rm var}(\hat{\gamma})}{(1+\gamma)^{4}},
\end{eqnarray}
where, again, $\gamma=\gamma_{*}$, the limit of $\hat{\gamma}$. Thus, an estimator of ${\rm var}(\hat{h}^{2})$ is given by
\begin{eqnarray}
\widehat{{\rm var}(\hat{h}^{2})}&=&\frac{\hat{C}}{(1+\hat{\gamma})^{4}\hat{B}^{2}}.
\end{eqnarray}

Note that all of the variance estimators obtained here are guaranteed to be nonnegative (and positive with probability one), a desirable property for a variance estimator. In particular, one can take square root of the variance estimator, and use it to construct a large-sample confidence interval for the corresponding parameter. Let $\theta$
denote a parameter of interest, such as $\sigma_{\epsilon}^{2}, h^{2}$, and $\hat{\theta}$ be its estimator. Let $\widehat{{\rm var}(\hat{\theta})}$ be a variance estimator
for $\hat{\theta}$ that is guaranteed nonnegative. Given $\alpha\in(0,1)$, by Theorem 2 of Jiang {\it et al.} (2016), an approximate $100(1-\alpha)\%$ confidence
interval for $\theta$ is given by
\begin{eqnarray}
\left[\hat{\theta}-z_{\alpha/2}\sqrt{\widehat{{\rm var}(\hat{\theta})}},\;\;\;\hat{\theta}+z_{\alpha/2}\sqrt{\widehat{{\rm var}(\hat{\theta})}}\right],
\end{eqnarray}
where $z_{\alpha/2}$ is the $\alpha/2$ critical value of $N(0,1)$ [i.e., ${\rm P}(Z>z_{\alpha/2})=\alpha/2$ for $Z\sim N(0,1)$].
\section{Simulation Studies}

We simulate scenarios similar to that in Jiang  {\it et al.} (2016). Specifically, we simulate the allele frequencies for $p$ SNPs from the Uniform[0.05, 0.5] distribution, and denote $f_{j}$ as the allele frequency of the $j$th SNP, for $j = 1,2, \dots p$. The genotype matrix $U \in \lbrace 0,1,2 \rbrace ^{n \times p}$ has rows corresponding to the individuals and the columns corresponding to the SNPs. The genotype value of each individual for the $j$th SNP is sampled from $\lbrace 0,1,2 \rbrace$ with probabilities $(1-f_{j})^2, 2f_{j}(1-f_{j}), f_{j}^2$, respectively. Let the standardized genotype matrix $Z$ be such that each column of $U$ is standardized to have zero mean and unit variance, and then let $\tilde{Z} = p^{-\frac{1}{2}}Z$. We express the relationship between the phenotypic vector $y$ and the standardized genotype matrix $\tilde{Z}$ in the LMM in (1).

As previously noted, (1) assumes that $\alpha_{j} \sim N(0, \sigma^2_{\alpha})$ for all $j \in \lbrace 1,2, \dots, p \rbrace$ when in reality, only a subset $m$ of the SNPs is associated with the phenotype. Thus, a correct model is (2) and the heritability should be

\begin{equation}
h_{\text{true}}^2 = \frac{(m/p)\sigma_{\alpha}^2}{(m/p)\sigma_{\alpha}^2 + \sigma_{\epsilon}^2}
\end{equation}

Since it is not possible to identify all of the $m$ SNPs in practice, we follow model (2) to simulate the phenotypes and use all of the SNPs in $Z$ to estimate the variance components, $\sigma_{\alpha}^2$ and $\sigma_{\epsilon}^2$, in model (1). We therefore estimate the heritability as

\begin{equation}
\hat{h}^2 = \frac{\hat{\sigma}_{\alpha}^2}{\hat{\sigma}_{\alpha}^2 + \hat{\sigma}_{\epsilon}^2},
\end{equation}
where the estimates of the variance components are their REML estimates.

In the simulations, given $n, p$, and $m$, we set the true parameters as $\mu = 0$, $\sigma^2_{\epsilon} = a$, and $\sigma^2_{\alpha} = b \frac{p}{m}$, for $(a,b) \in \lbrace (0.8, 0.2), (0.6, 0.4), (0.4, 0.6), (0.2, 0.8) \rbrace$, where the heritability parameter is varied. We performed simulations with those true parameters under misspecifications of $\omega \in \lbrace 0.005, 0.01, 0.05, 0.1, 0.5 \rbrace$. Note that $\gamma = \dfrac{\sigma^2_\alpha}{\sigma^2_\epsilon}$. We simulated the data under model (2), but found REML estimates under model (1). For each scenario, we carried out $300$ replications, and report the results (see below).

Let ${\rm var}(\hat{\theta})$ be the sample variance of all of the simulated $\hat{\theta}$s for $\theta \in \lbrace \sigma^2_{\epsilon}, h^2 \rbrace$, $\hat{v} = \widehat{{\rm var}(\hat{\theta})}$, and ${\rm E}(\hat{v})$ be the sample mean of all of the simulated $\hat{v}$s. The percentage of risk bias (\%RB) is defined as
$$\%{\rm RB}=100\times\left\{\dfrac{{\rm E}(\hat{v})-{\rm var}(\hat{\theta})}{{\rm var}(\hat{\theta})}\right\}.$$
We also look at the sample standard deviation of all of the simulated $\hat{v}$s, denoted as s($\hat{v}$). The N$_\lambda$ for $\lambda \in \lbrace 0.01, 0.05, 0.1 \rbrace$ is the empirical coverage probability for large sample confidence intervals of $\theta$ at level $\lambda$. Since $\theta$ is bounded, we also consider the large sample truncated confidence intervals of $\theta$ and denote the empirical coverage probability by T$_\lambda$ for $\lambda \in \lbrace 0.01, 0.05, 0.1 \rbrace$. To find the truncated confidence intervals of $\theta \in \left\lbrace \sigma^2_{\epsilon}, h^2 \right\rbrace$, we use the quantiles of the truncated normal distribution, where the mean is the REML estimate of $\theta$ and the variance is the variance estimate of REML estimate of $\theta$. In particular, since the lower bound of $\theta = \sigma_{\epsilon}^2$ is $0$, we truncate the lower bound by $0$ but not the upper bound. For $\theta = h^2$, we truncate the lower bound by $0$ and the upper bound by $1$. The quantiles of the truncated normal distribution at levels $\lambda/2$ and $1 - \lambda/2$ are the lower and upper bounds of the confidence interval, respectively.

Tables 1 and 2 showcase some results for $\theta = \sigma^2_{\epsilon}$, and Tables 3 and 4 show a few for $\theta = h^2$. Other simulations are given in the supplementary materials.

\begin{table}[H] \centering 
	\caption{$\sigma^2_{\epsilon_0} = 0.8, \sigma^2_{\alpha_0} = 0.2\dfrac{p}{m}$ for $\theta = \sigma^2_{\epsilon}$, $(n,p) = (2000, 20000)$} 
	\label{} 
	\begin{tabular}{ccccc|ccc|ccc} 
		\\[-2.8ex]\hline 
		\hline \\[-3.8ex] 
		$m$ & \% RB & ${\rm var}(\hat{\theta})$ & ${\rm E}(\hat{v})$ & ${\rm s}(\hat{v})$ & N$_{0.01}$ & N$_{0.05}$ & N$_{0.1}$ & T$_{0.01}$ & T$_{0.05}$ & T$_{0.1}$ \\ 
		\hline \\[-3.8ex] 
		$20$ & $5.442$ & $0.010$ & $0.011$ & $0.001$ & $0.990$ & $0.970$ & $0.910$ & $0.990$ & $0.970$ & $0.910$ \\
		$200$ & $$-$1.970$ & $0.011$ & $0.010$ & $0.001$ & $0.987$ & $0.940$ & $0.887$ & $0.987$ & $0.940$ & $0.887$ \\
		$2,000$ & $12.302$ & $0.009$ & $0.010$ & $0.001$ & $0.997$ & $0.960$ & $0.900$ & $0.997$ & $0.960$ & $0.900$ \\
		$20,000$ & $3.209$ & $0.010$ & $0.011$ & $0.001$ & $0.993$ & $0.947$ & $0.903$ & $0.993$ & $0.947$ & $0.903$ \\
		\hline \\[-3.8ex]
		\end{tabular}
\end{table}


\begin{table}[H] \centering 
	\caption{$\sigma^2_{\epsilon_0} = 0.4, \sigma^2_{\alpha_0} = 0.6\dfrac{p}{m}$ for $\theta = \sigma^2_{\epsilon}$, $\omega = m/p = 0.1$} 
	\label{} 
	\begin{tabular}{ccccc|ccc|ccc}  
		\\[-2.8ex]\hline 
		\hline \\[-3.8ex] 
		$(n,p)$ & \% RB & ${\rm var}(\hat{\theta})$ & ${\rm E}(\hat{v})$ & ${\rm s}(\hat{v})$ & N$_{0.01}$ & N$_{0.05}$ & N$_{0.1}$ & T$_{0.01}$ & T$_{0.05}$ & T$_{0.1}$ \\ 
		\hline \\[-3.8ex]
		
		(1000, 10000) & $0.828$ & $0.018$ & $0.018$ & $0.002$ & $0.990$ & $0.933$ & $0.893$  & $0.993$ & $0.937$ & $0.900$ \\ 
		(2000, 20000) & $7.608$ & $0.008$ & $0.009$ & $0.001$ & $0.990$ & $0.953$ & $0.907$  & $0.990$ & $0.957$ & $0.907$ \\ 
		(3000, 30000) & $-6.494$ & $0.007$ & $0.006$ & $0.0004$ & $0.980$ & $0.940$ & $0.897$ & $0.980$ & $0.940$ & $0.897$ \\ 
		(4000, 40000) & $-6.731$ & $0.005$ & $0.005$ & $0.0003$ & $0.987$ & $0.947$ & $0.887$ & $0.987$ & $0.947$ & $0.887$ \\ 
		(5000, 50000) & $1.433$ & $0.004$ & $0.004$ & $0.0002$ & $0.993$ & $0.957$ & $0.893$  & $0.993$ & $0.957$ & $0.893$ \\ 
		\hline \\[-3.8ex] 
	\end{tabular} 
\end{table}

\begin{table}[H] \centering 
	\caption{$\sigma^2_{\epsilon_0} = 0.8, \sigma^2_{\alpha_0} = 0.2\dfrac{p}{m}$ for $\theta = h^2$, $\omega = m/p = 0.05$} 
	\label{} 
	\begin{tabular}{ccccc|ccc|ccc}  
		\\[-2.8ex]\hline 
		\hline \\[-3.8ex]
		$(n,p)$ & \% RB & ${\rm var}(\hat{\theta})$ & ${\rm E}(\hat{v})$ & ${\rm s}(\hat{v})$ & N$_{0.01}$ & N$_{0.05}$ & N$_{0.1}$ & T$_{0.01}$ & T$_{0.05}$ & T$_{0.1}$ \\ 
		\hline \\[-3.8ex]  
		
		(1000, 10000) & $20.046$ & $0.017$ & $0.020$ & $0.0002$ & $0.990$ & $0.977$ & $0.960$ & $0.990$ & $0.977$ & $0.960$ \\ 
		(2000, 20000) & $2.164$ & $0.010$ & $0.010$ & $0.0001$ & $0.993$ & $0.960$ & $0.890$ & $0.993$ & $0.980$ & $0.920$ \\ 
		(3000, 30000) & $$-$7.961$ & $0.007$ & $0.007$ & $0.00004$ & $0.987$ & $0.933$ & $0.880$ & $0.987$ & $0.947$ & $0.893$ \\ 
		(4000, 40000) & $3.595$ & $0.005$ & $0.005$ & $0.00002$ & $0.993$ & $0.947$ & $0.907$ & $0.993$ & $0.953$ & $0.917$ \\ 
		(5000, 50000) & $$-$8.444$ & $0.004$ & $0.004$ & $0.00001$ & $0.983$ & $0.937$ & $0.887$ & $0.987$ & $0.937$ & $0.887$ \\ 
		\hline \\[-3.8ex]
	\end{tabular}
\end{table}

\begin{table}[H] \centering 
	\caption{$\sigma^2_{\epsilon_0} = 0.4, \sigma^2_{\alpha_0} = 0.6\dfrac{p}{m}$ for $\theta = h^2$, $\omega = m/p = 0.1$} 
	\label{} 
	\begin{tabular}{ccccc|ccc|ccc}  
		\\[-2.8ex]\hline 
		\hline \\[-3.8ex] 
		$(n,p)$ & \% RB & ${\rm var}(\hat{\theta})$ & ${\rm E}(\hat{v})$ & ${\rm s}(\hat{v})$ & N$_{0.01}$ & N$_{0.05}$ & N$_{0.1}$ & T$_{0.01}$ & T$_{0.05}$ & T$_{0.1}$ \\ 
		\hline \\[-3.8ex]
		
		(1000, 10000) & $-2.578$ & $0.019$ & $0.019$ & $0.001$ & $0.983$ & $0.923$ & $0.887$ & $0.990$ & $0.937$ & $0.890$ \\ 
		(2000, 20000) & $4.057$ & $0.009$ & $0.009$ & $0.0004$ & $0.990$ & $0.953$ & $0.897$ & $0.990$ & $0.953$ & $0.900$ \\ 
		(3000, 30000) & $-8.355$ & $0.007$ & $0.006$ & $0.0002$ & $0.983$ & $0.930$ & $0.893$ & $0.983$ & $0.930$ & $0.893$ \\ 
		(4000, 40000) & $-4.804$ & $0.005$ & $0.005$ & $0.0001$ & $0.990$ & $0.943$ & $0.890$ & $0.990$ & $0.943$ & $0.890$ \\ 
		(5000, 50000) & $-0.360$ & $0.004$ & $0.004$ & $0.0001$ & $0.990$ & $0.947$ & $0.897$ & $0.990$ & $0.947$ & $0.897$ \\ 
		\hline \\[-3.8ex] 
	\end{tabular} 
\end{table}

We can compare the results of our method with GCTA, where the comparisons occur between Tables 5 and 6, Tables 7 and 8, Tables 9 and 10, and Tables 11 and 12. Note that our method and GCTA perform similarly well.

\begin{table}[H] \centering 
	\caption{$\sigma^2_{\epsilon_0} = 0.4, \sigma^2_{\alpha_0} = 0.6\dfrac{p}{m}$ for $\theta = h^2$, $(n,p) = (2000, 20000)$} 
	\label{} 
	\begin{tabular}{ccccc|ccc|ccc}  
		\\[-2.8ex]\hline 
		\hline \\[-3.8ex] 
		$m$ & \% RB & ${\rm var}(\hat{\theta})$ & ${\rm E}(\hat{v})$ & ${\rm s}(\hat{v})$ & N$_{0.01}$ & N$_{0.05}$ & N$_{0.1}$  & T$_{0.01}$ & T$_{0.05}$ & T$_{0.1}$ \\ 
		\hline \\[-3.8ex] 
		$20$ & $3.208$ & $0.009$ & $0.010$ & $0.003$ & $0.983$ & $0.947$ & $0.900$ & $0.983$ & $0.947$ & $0.900$ \\ 
		$200$ & $$-$10.250$ & $0.010$ & $0.009$ & $0.001$ & $0.983$ & $0.930$ & $0.860$ & $0.983$ & $0.930$ & $0.860$ \\ 
		$2,000$ & $7.608$ & $0.008$ & $0.009$ & $0.001$ & $0.990$ & $0.953$ & $0.907$ & $0.990$ & $0.957$ & $0.907$ \\ 
		$20,000$ & $$-$3.961$ & $0.010$ & $0.009$ & $0.001$ & $0.990$ & $0.947$ & $0.890$ & $0.990$ & $0.947$ & $0.890$ \\  
		\hline \\[-3.8ex] 
	\end{tabular} 
\end{table}

\begin{table}[H] \centering 
	\caption{GCTA CIs: $\sigma^2_{\epsilon_0} = 0.4, \sigma^2_{\alpha_0} = 0.6\dfrac{p}{m}$ for $\theta = \sigma^2_{\epsilon}$, $(n,p) = (2000, 20000)$} 
	\label{} 
	\begin{tabular}{ccccc|ccc} 
		\\[-2.8ex]\hline 
		\hline \\[-3.8ex] 
		$m$ & \% RB & ${\rm var}(\hat{\theta})$ & ${\rm E}(\hat{v})$ & ${\rm s}(\hat{v})$ & GCTA$_{0.01}$ & GCTA$_{0.05}$ & GCTA$_{0.1}$ \\ 
		\hline \\[-3.8ex] 
		$20$ &  3.3062 & 0.0092 & 0.0095 & 0.0035 & 0.9867 & 0.9467 & 0.9067 \\ 
		$200$ &  -9.9262 & 0.0102 & 0.0092 & 0.0013 & 0.9833 & 0.93 & 0.8633\\ 
		$2,000$ & 7.6525 & 0.0085 & 0.0091 & 8e-04 & 0.99 & 0.95 & 0.91  \\ 
		$20,000$ &  -3.7719 & 0.0095 & 0.0091 & 8e-04 & 0.99 & 0.9433 & 0.8933  \\ 
		\hline \\[-3.8ex] 
	\end{tabular} 
\end{table}

\begin{table}[H] \centering 
\caption{$\sigma^2_{\epsilon_0} = 0.4, \sigma^2_{\alpha_0} = 0.6\dfrac{p}{m}$ for $\theta = h^2$, $(n,p) = (2000, 20000)$} 
\label{} 
\begin{tabular}{ccccc|ccc|ccc}  
\\[-2.8ex]\hline 
\hline \\[-3.8ex] 
$m$ & \% RB & ${\rm var}(\hat{\theta})$ & ${\rm E}(\hat{v})$ & ${\rm s}(\hat{v})$ & N$_{0.01}$ & N$_{0.05}$ & N$_{0.1}$ & T$_{0.01}$ & T$_{0.05}$ & T$_{0.1}$ \\ 
\hline \\[-3.8ex] 
$20$ & $3.208$ & $0.009$ & $0.010$ & $0.003$ & $0.983$ & $0.947$ & $0.900$ & $0.983$ & $0.947$ & $0.900$ \\ 
$200$ & $$-$10.250$ & $0.010$ & $0.009$ & $0.001$ & $0.983$ & $0.930$ & $0.860$ & $0.983$ & $0.930$ & $0.860$ \\ 
$2,000$ & $7.608$ & $0.008$ & $0.009$ & $0.001$ & $0.990$ & $0.953$ & $0.907$ & $0.990$ & $0.957$ & $0.907$ \\ 
$20,000$ & $$-$3.961$ & $0.010$ & $0.009$ & $0.001$ & $0.990$ & $0.947$ & $0.890$ & $0.990$ & $0.947$ & $0.890$ \\  
\hline \\[-3.8ex] 
\end{tabular} 
\end{table} 

\begin{table}[H] \centering 
	\caption{GCTA CIs: $\sigma^2_{\epsilon_0} = 0.4, \sigma^2_{\alpha_0} = 0.6\dfrac{p}{m}$ for $\theta = h^2$, $(n,p) = (2000, 20000)$} 
	\label{} 
	\begin{tabular}{ccccc|ccc} 
		\\[-2.8ex]\hline 
		\hline \\[-3.8ex] 
		$m$ & \% RB & ${\rm var}(\hat{\theta})$ & ${\rm E}(\hat{v})$ & ${\rm s}(\hat{v})$ & GCTA$_{0.01}$ & GCTA$_{0.05}$ & GCTA$_{0.1}$ \\ 
		\hline \\[-3.8ex] 
		$20$ & -33.6238 & 0.0142 & 0.0094 & 5e-04 & 0.95 & 0.8833 & 0.8133 \\ 
		$200$ &  -15.1985 & 0.0111  & 0.0094 & 5e-04 & 0.9767 & 0.9267 & 0.8467\\ 
		$2,000$ &  4.0708 & 0.009 & 0.0094 & 5e-04 & 0.9867 & 0.9567 & 0.9\\ 
		$20,000$ & -1.8037 & 0.0096 & 0.0094 & 5e-04 & 0.99 & 0.9467 & 0.9033  \\ 
		\hline \\[-3.8ex] 
	\end{tabular} 
\end{table}

\begin{table}[H] \centering 
\caption{$\sigma^2_{\epsilon_0} = 0.4, \sigma^2_{\alpha_0} = 0.6\dfrac{p}{m}$ for $\theta = \sigma^2_{\epsilon}$, $\omega = m/p = 0.01$} 
\label{} 
\begin{tabular}{ccccc|ccc|ccc}  
\\[-2.8ex]\hline 
\hline \\[-3.8ex] 
$(n,p)$ & \% RB & ${\rm var}(\hat{\theta})$ & ${\rm E}(\hat{v})$ & ${\rm s}(\hat{v})$ & N$_{0.01}$ & N$_{0.05}$ & N$_{0.1}$ & T$_{0.01}$ & T$_{0.05}$ & T$_{0.1}$ \\ 
\hline \\[-3.8ex] 
$(1000, 10000)$ & $-8.854$ & $0.020$ & $0.019$ & $0.003$ & $0.990$ & $0.927$ & $0.877$  & $0.990$ & $0.940$ & $0.883$ \\ 
$(2000, 20000)$ & $$-$10.250$ & $0.010$ & $0.009$ & $0.001$ & $0.983$ & $0.930$ & $0.860$ & $0.983$ & $0.930$ & $0.860$ \\ 
$(3000, 30000)$ & $5.450$ & $0.006$ & $0.006$ & $0.001$ & $0.997$ & $0.957$ & $0.907$ & $0.997$ & $0.957$ & $0.907$ \\ 
$(4000, 40000)$ & $$-$16.138$ & $0.005$ & $0.005$ & $0.0004$ & $0.983$ & $0.917$ & $0.870$ & $0.983$ & $0.917$ & $0.870$ \\ 
$(5000, 50000)$ & $10.734$ & $0.003$ & $0.004$ & $0.0003$ & $0.987$ & $0.967$ & $0.923$ & $0.987$ & $0.967$ & $0.923$ \\ 
\hline \\[-3.8ex] 
\end{tabular} 
\end{table}

\begin{table}[H] \centering 
	\caption{GCTA CIs: $\sigma^2_{\epsilon_0} = 0.4, \sigma^2_{\alpha_0} = 0.6\dfrac{p}{m}$ for $\theta = \sigma^2_{\epsilon}$, $\omega = m/p = 0.01$} 
	\label{} 
	\begin{tabular}{ccccc|ccc}  
		\\[-2.8ex]\hline 
		\hline \\[-3.8ex] 
		$(n,p)$ & \% RB & ${\rm var}(\hat{\theta})$ & ${\rm E}(\hat{v})$ & ${\rm s}(\hat{v})$ & GCTA$_{0.01}$ & GCTA$_{0.05}$ & GCTA$_{0.1}$ \\ 
		\hline \\[-3.8ex] 
		$(1000, 10000)$ & -8.4776 & 0.0204 & 0.0186 & 0.0035 & 0.9933 & 0.93 & 0.8767 \\ 
		$(2000, 20000)$ & -9.9262 & 0.0102 & 0.0092 & 0.0013 & 0.9833 & 0.93 & 0.8633\\ 
		$(3000, 30000)$ &  5.4081 & 0.0058 & 0.0061 & 7e-04 & 1 & 0.9567 & 0.9067 \\ 
		$(4000, 40000)$ & -16.1722 & 0.0055 & 0.0046 & 4e-04 & 0.9867 & 0.92 & 0.8667 \\ 
		$(5000, 50000)$ &  10.8681 & 0.0033 & 0.0037 & 3e-04 & 0.9867 & 0.9633 & 0.9233 \\ 
		\hline \\[-3.8ex] 
	\end{tabular} 
\end{table} 

\begin{table}[H] \centering 
\caption{$\sigma^2_{\epsilon_0} = 0.4, \sigma^2_{\alpha_0} = 0.6\dfrac{p}{m}$ for $\theta = h^2$, $\omega = m/p = 0.01$} 
\label{} 
\begin{tabular}{ccccc|ccc|ccc}  
\\[-2.8ex]\hline 
\hline \\[-3.8ex] 
$(n,p)$ & \% RB & ${\rm var}(\hat{\theta})$ & ${\rm E}(\hat{v})$ & ${\rm s}(\hat{v})$ & N$_{0.01}$ & N$_{0.05}$ & N$_{0.1}$ & T$_{0.01}$ & T$_{0.05}$ & T$_{0.1}$ \\ 
\hline \\[-3.8ex] 
$(1000, 10000)$ & $$-$12.968$ & $0.022$ & $0.019$ & $0.001$ & $0.987$ & $0.913$ & $0.873$ & $0.990$ & $0.917$ & $0.877$ \\ 
$(2000, 20000)$ & $$-$15.510$ & $0.011$ & $0.009$ & $0.0004$ & $0.980$ & $0.927$ & $0.850$ & $0.980$ & $0.927$ & $0.850$ \\ 
$(3000, 30000)$ & $$-$2.363$ & $0.006$ & $0.006$ & $0.0002$ & $0.993$ & $0.957$ & $0.893$ & $0.993$ & $0.957$ & $0.893$ \\ 
$(4000, 40000)$ & $$-$22.084$ & $0.006$ & $0.005$ & $0.0001$ & $0.983$ & $0.910$ & $0.860$ & $0.983$ & $0.910$ & $0.860$ \\ 
$(5000, 50000)$ & $5.989$ & $0.004$ & $0.004$ & $0.0001$ & $0.993$ & $0.950$ & $0.920$ & $0.993$ & $0.950$ & $0.920$ \\ 
\hline \\[-3.8ex] 
\end{tabular} 
\end{table}

\begin{table}[H] \centering 
	\caption{GCTA CIs: $\sigma^2_{\epsilon_0} = 0.4, \sigma^2_{\alpha_0} = 0.6\dfrac{p}{m}$ for $\theta = h^2$, $\omega = m/p = 0.01$} 
	\label{} 
	\begin{tabular}{ccccc|ccc}  
		\\[-2.8ex]\hline 
		\hline \\[-3.8ex] 
		$(n,p)$ & \% RB & ${\rm var}(\hat{\theta})$ & ${\rm E}(\hat{v})$ & ${\rm s}(\hat{v})$ & GCTA$_{0.01}$ & GCTA$_{0.05}$ & GCTA$_{0.1}$ \\ 
		\hline \\[-3.8ex] 
		$(1000, 10000)$ &  -12.6111 & 0.0216 & 0.0189 & 0.0014 & 0.99 & 0.91 & 0.87 \\ 
		$(2000, 20000)$ & -15.1985 & 0.0111  & 0.0094 & 5e-04 & 0.9767 & 0.9267 & 0.8467 \\ 
		$(3000, 30000)$ & -2.3934 & 0.0065 & 0.0063 & 3e-04 & 0.9933 & 0.9567 & 0.8933 \\ 
		$(4000, 40000)$ &  -22.1093 & 0.0061 & 0.0047 & 2e-04 & 0.9833 & 0.91 & 0.8567 \\ 
		$(5000, 50000)$ & 6.1401 & 0.0036 & 0.0038 & 1e-04 & 0.9933 & 0.95 & 0.9233 \\ 
		\hline \\[-3.8ex] 
	\end{tabular} 
\end{table} 

{\bf Remark:} The GCTA method is based on the standard LMM analysis (e.g., Jiang 2007), which does not take into account (i) that the LMM is misspecified (see Section 1); (ii) that the design matrix, $Z$, for the random effects is random; and (iii) the asymptotic framework is different than the standard assumption that the number of random effects is, at most, of the same order as the sample size. In typical GWAS, the number of random effects, which correspond to the SNPs, is typically of higher order than the sample size. On the other hand, our method is fully supported by the recently established theory on high-dimensional mis-LMM analysis (Jiang {\it et al.} 2016), based on which, the variance estimators are derived in the current paper. Thus, in a way, the results of this paper have provided justification for the use of the GCTA method for inference.

\section{Real data example}
We applied this method to a real data example using a subset of indviduals with height data from the UK Biobank (UKBB) database. We consider $n=$ 4,986 Caucasian individuals who are unrelated up to the 3rd degree using KING (Manichaikul {\it et al.} 2010) to avoid inflating the heritability estimate. The UKBB performed genotype imputation using IMPUTE4 and the Haplotype Reference Consortium reference panel (Bycroft {\it et al.} 2018). We retained imputed SNPs with INFO scores greater than 0.8. Then, we removed imputed SNPs with a missing call rate exceeding 0.05, a Hardy-Weinberg equilibrium exact test p-value below $1\times10^{-10}$, or a minor allele frequency below 0.05. After quality control, $p=$ 6,133,110 SNPs remained for analysis.

Then, we applied the LMM approach described in model (1) to obtain REML estimates of the variance components, and then estimated their variances and construct confidence intervals for the parameters of interest. For the matrix $X$ of fixed effects, in addition to the intercept, we accounted for sex, age, and population stratification using the first twenty principal component scores derived from genotype data provided by the UKBB.

We obtained REML estimates $\hat{\sigma}^2_{\epsilon} = 20.150$, $\hat{\gamma} = 1.003$, and $\hat{h}^2 = 0.5009$. Using our approach, we got the following variance estimates for our parameter of interests: $\widehat{{\rm var}(\hat{\sigma}_{\epsilon}^{2})} = 7.117$, and  $\widehat{{\rm var}(\hat{h}^{2})} = 0.0045$. The variance estimates from GCTA are comparable:  $\widehat{{\rm var}(\hat{\sigma}_{\epsilon}^{2})} = 7.242$, $\widehat{{\rm var}(\hat{h}^{2})} = 0.0046$. The corresponding $95 \%$ confidence intervals for $\sigma^2_{\epsilon}$ are (14.921, 25.379) and (14.875, 25.424) for our method and GCTA, respectively. The $95 \%$ confidence intervals for $h^2$ are (0.3697, 0.6321) and (0.3685, 0.6332) for our method and GCTA, respectively. The heritability of height estimated by LMM/REML have similar results in other data sets (e.g., Yang {\it et al.}(2010), Zhou {\it et al.} (2013), Golan {\it et al.} (2014)).

\section*{Appendix}
\subsection*{A.1\hspace{2mm}Derivation of (4)}
\hspace{4mm}
Using the identity $B^{-1}=A^{-1}+A^{-1}(A-B)B^{-1}$, the following first-order approximations can be derived: $P_{\hat{\gamma}}\approx P_{\gamma}-(\hat{\gamma}-\gamma)Q_{\gamma}$, where $\gamma=\gamma_{*}$ which is the limit of $\hat{\gamma}$, and $Q_{\gamma}=P_{\gamma}\tilde{Z}\tilde{Z}'P_{\gamma}$. With those,
and (3), the following approximation can be derived:
$$\hat{\sigma}_{\epsilon}^{2}\approx\frac{y'P_{\gamma}^{2}y}{{\rm
tr}(P_{\gamma})}-(\hat{\gamma}-\gamma)S_{\gamma,y},\eqno{\rm (A.1)}$$
where
$$S_{\gamma,y}=\frac{y'R_{\gamma}y}{{\rm tr}(P_{\gamma})}-\frac{{\rm
tr}(Q_{\gamma})}{{\rm tr}^{2}(P_{\gamma})}y'P_{\gamma}^{2}y\eqno{\rm (A.2)}$$
with $R_{\gamma}=P_{\gamma}Q_{\gamma}+Q_{\gamma}P_{\gamma}$.

Next, we obtain an expansion for $\hat{\gamma}-\gamma$. From (3) of Jiang {\it et al.} (2016), we have
$$\frac{yP_{\gamma}\tilde{Z}\tilde{Z}'P_{\gamma}y}{{\rm tr}(P_{\gamma}\tilde{Z}\tilde{Z}')}=\frac{yP_{\gamma}^{2}y}{{\rm tr}(P_{\gamma})}.\eqno{\rm (A.3)}$$
The RHS (righthand side) of (A.3) is approximated by (A.1). As for the LHS (lefthand size) of (A.3), one can derive $Q_{\hat{\gamma}}\approx Q_{\gamma}-(\hat
{\gamma}-\gamma)T_{\gamma}$, where $T_{\gamma}=P_{\gamma}\tilde{Z}\tilde{Z}'Q_{\gamma}+Q_{\gamma}\tilde{Z}\tilde{Z}'P_{\gamma}$. Furthermore,
using the elementary expansion of Jiang 2010 (p. 103), we have
$$\frac{1}{{\rm tr}(P_{\hat{\gamma}}\tilde{Z}\tilde{Z}')}\approx\frac
{1}{{\rm tr}(P_{\gamma}\tilde{Z}\tilde{Z}')}+\frac{(\hat{\gamma}-\gamma){\rm
tr}(Q_{\gamma}\tilde{Z}\tilde{Z}')}{{\rm tr}^{2}(P_{\gamma}\tilde{Z}\tilde
{Z}')}.$$
Thus, the LHS of (A.3) can be approximated by
\begin{eqnarray*}
&&\{y'Q_{\gamma}y-(\hat{\gamma}-\gamma)y'T_{\gamma}y\}\left\{\frac{1}{{\rm
 tr}(P_{\gamma}\tilde{Z}\tilde{Z}')}+\frac{(\hat{\gamma}-\gamma){\rm
tr}(Q_{\gamma}\tilde{Z}\tilde{Z}')}{{\rm tr}^{2}(P_{\gamma}\tilde{Z}\tilde
{Z}')}\right\}\nonumber\\
&\approx&\frac{y'Q_{\gamma}y}{{\rm tr}(P_{\gamma}\tilde{Z}\tilde{Z}')}-(\hat
{\gamma}-\gamma)U_{\gamma,y},\hspace{86mm}{\rm (A.4)}
\end{eqnarray*}
where
$$U_{\gamma,y}=\frac{y'T_{\gamma}y}{{\rm tr}(P_{\gamma}\tilde{Z}\tilde{Z}')}
-\frac{{\rm tr}(Q_{\gamma}\tilde{Z}\tilde{Z}')}{{\rm tr}^{2}(P_{\gamma}\tilde
{Z}\tilde{Z}')}y'Q_{\gamma}y.$$
By equating the LHS to the RHS, i.e., (A.1) to (A.4), we obtain the following:
\begin{eqnarray*}
\hat{\gamma}-\gamma&\approx&\frac{1}{S_{\gamma,y}-U_{\gamma,y}}\left\{\frac
{y'P_{\gamma}^{2}y}{{\rm tr}(P_{\gamma})}-\frac{y'Q_{\gamma}y}{{\rm
tr}(P_{\gamma}\tilde{Z}\tilde{Z}')}\right\}\nonumber\\
&\approx&\frac{1}{{\rm E}(S_{\gamma,y})-{\rm E}(U_{\gamma,y})}\left\{\frac
{y'P_{\gamma}^{2}y}{{\rm tr}(P_{\gamma})}-\frac{y'Q_{\gamma}y}{{\rm
tr}(P_{\gamma}\tilde{Z}\tilde{Z}')}\right\}.\hspace{45mm}{\rm (A.5)}
\end{eqnarray*}
Combining (A.1) and (A.5), we obtain
\begin{eqnarray*}
\hat{\sigma}_{\epsilon}^{2}&\approx&\frac{y'P_{\gamma}^{2}y}{{\rm
tr}(P_{\gamma})}+\frac{S_{\gamma,y}}{{\rm E}(U_{\gamma,y})-{\rm
E}(S_{\gamma,y})}\left\{\frac{y'P_{\gamma}^{2}y}{{\rm tr}(P_{\gamma})}
-\frac{y'Q_{\gamma}y}{{\rm tr}(P_{\gamma}\tilde{Z}\tilde{Z}')}\right\}
\nonumber\\
&\approx&\frac{y'P_{\gamma}^{2}y}{{\rm tr}(P_{\gamma})}+\frac{{\rm
E}(S_{\gamma,y})}{{\rm E}(U_{\gamma,y})-{\rm
E}(S_{\gamma,y})}\left\{\frac{y'P_{\gamma}^{2}y}{{\rm tr}(P_{\gamma})}
-\frac{y'Q_{\gamma}y}{{\rm tr}(P_{\gamma}\tilde{Z}\tilde{Z}')}\right\}
\nonumber\\
&=&\frac{{\rm E}(U_{\gamma,y})}{{\rm E}(U_{\gamma,y})-{\rm
E}(S_{\gamma,y})}\cdot\frac{y'P_{\gamma}^{2}y}{{\rm tr}(P_{\gamma})}
\nonumber\\
&&+\frac{{\rm E}(U_{\gamma,y})}{{\rm E}(S_{\gamma,y})-{\rm
E}(U_{\gamma,y})}\cdot\frac{y'Q_{\gamma}y}{{\rm tr}(P_{\gamma}\tilde
{Z}\tilde{Z}')}\nonumber\\
&=&\tilde{\sigma}_{\epsilon}^{2}.
\end{eqnarray*}
\subsection*{A.2\hspace{2mm}Further approximation to the right side of (6)}
\hspace{4mm}
Note that $\tilde{\sigma}_{\epsilon}^{2}=y'D_{\gamma}y$ for some matrix
$D_{\gamma}$. Thus, by the normal theory (e.g., Jiang 2007, p. 238), we have
$${\rm var}(\tilde{\sigma}_{\epsilon}^{2}|Z)=2{\rm tr}(D_{\gamma}\Sigma
D_{\gamma}\Sigma)=2{\rm tr}\left\{\left(\sigma_{\epsilon}^{2}D_{\gamma}
+\sigma_{\alpha}^{2}\sum_{i=1}^{m}D_{\gamma}\tilde{Z}_{i}\tilde{Z}_{i}'
\right)^{2}\right\},$$
where $\Sigma=\sigma_{\epsilon}^{2}I_{n}+\sigma_{\alpha}^{2}\sum_{i=1}^{m}
\tilde{Z}_{i}\tilde{Z}_{i}'$ is the true covariance matrix of $y$. Therefore,
we have ${\rm E}\{{\rm var}(\tilde{\sigma}_{\epsilon}^{2}|Z)\}=2\{
\sigma_{\epsilon}^{4}{\rm E}(D_{\gamma}^{2})+2\sigma_{\epsilon}^{2}
\sigma_{\alpha}^{2}{\rm E}(I_{1})+\sigma_{\alpha}^{4}{\rm E}(I_{2})\}$ with
$I_{1}=\sum_{i=1}^{m}{\rm tr}(D_{\gamma}\tilde{Z}_{i}\tilde{Z}_{i}'
D_{\gamma})$ and $I_{2}=\sum_{i,j=1}^{m}{\rm tr}(D_{\gamma}\tilde{Z}_{i}
\tilde{Z}_{i}'D_{\gamma}\tilde{Z}_{j}\tilde{Z}_{j}')$. By the fact that
$Z_{i}, 1\leq i\leq p$ are i.i.d., it can be shown that ${\rm E}(I_{1})=
\omega{\rm E}\{{\rm tr}(D_{\gamma}\tilde{Z}\tilde{Z}'D_{\gamma})\}$, and
${\rm E}(I_{2})\approx\omega^{2}{\rm E}\{{\rm tr}(D_{\gamma}\tilde{Z}
\tilde{Z}'D_{\gamma}\tilde{Z}\tilde{Z}')\}$, where $\omega=m/p$. It follows
that the RHS of (6) is approximately equal to
$$2{\rm E}\left[{\rm tr}\left\{(\sigma_{\epsilon}^{2}D_{\gamma}+\omega
\sigma_{\alpha}^{2}D_{\gamma}\tilde{Z}\tilde{Z}')^{2}\right\}\right]
=2\sigma_{\epsilon}^{4}{\rm E}\left[{\rm tr}\{(aP_{\gamma}+bP_{\gamma}
\tilde{Z}\tilde{Z}')^{2}\}\right],\eqno{\rm (A.6)}$$
where
$$a=\frac{{\rm E}(U_{\gamma,y})}{{\rm tr}(P_{\gamma})\{{\rm E}(U_{\gamma,y})
-{\rm E}(S_{\gamma,y})\}},\;\;\;b=\frac{{\rm E}(S_{\gamma,y})}{{\rm
tr}(P_{\gamma}\tilde{Z}\tilde{Z}')\{{\rm E}(S_{\gamma,y})-{\rm E}(U_{\gamma,
y})\}}.$$
Thus, we have
$$aP_{\gamma}+bP_{\gamma}\tilde{Z}\tilde{Z}'=\frac{1}{{\rm E}(U_{\gamma,
y})-{\rm E}(S_{\gamma,y})}\left\{\frac{{\rm E}(U_{\gamma,y})}{{\rm
tr}(P_{\gamma})}P_{\gamma}-\frac{{\rm E}(S_{\gamma,y})}{{\rm
tr}(P_{\gamma}\tilde{Z}\tilde{Z}')}P_{\gamma}\tilde{Z}\tilde{Z}'\right\}.
\eqno{\rm (A.7)}$$
Furthermore, using, once again, the i.i.d. property, it can be shown that
$${\rm E}(S_{\gamma,y})=\sigma_{\epsilon}^{2}{\rm E}\left\{\frac{{\rm
tr}(Q_{\gamma})}{{\rm tr}(P_{\gamma})}\right\},\;\;\;{\rm E}(U_{\gamma,
y})=\sigma_{\epsilon}^{2}{\rm E}\left\{\frac{{\rm tr}(Q_{\gamma}\tilde
{Z}\tilde{Z}')}{{\rm tr}(P_{\gamma}\tilde{Z}\tilde{Z}')}\right\}.\eqno{\rm (A.8)}$$
Combining (6), (A.6)--(A.8), it can be shown ${\rm var}(\hat{\sigma}_{\epsilon}^{2})
\approx 2\sigma_{\epsilon}^{2}{\rm E}(A)/\{{\rm E}(B)\}^{2}$, where $A, B$ are
given below (6).

\end{document}